\documentclass{ws-ijgmmp}

\usepackage{ tipa }
\usepackage{mathtools}
\usepackage{amsmath}
\usepackage{stmaryrd}
\usepackage{dsfont}
\usepackage{verbatim}
\usepackage{tikz}
\usetikzlibrary{shapes.geometric}
\usepackage{tikz-3dplot}
\usetikzlibrary{calc}
\usetikzlibrary{decorations.markings}

\newcommand{\be}{\begin{equation}}
\newcommand{\ee}{\end{equation}}
\newcommand{\ba}{\begin{eqnarray}}
\newcommand{\ea}{\end{eqnarray}}
\newcommand{\lp}{\left(}
\newcommand{\rp}{\right)}
\newcommand{\lb}{\left[}
\newcommand{\rb}{\right]}

\newcommand{\bbe}{\boldsymbol{\rm e}}
\newcommand{\bbie}{\boldsymbol{\textbf{\textschwa}}}
\newcommand{\ie}{\text{\textschwa}}
\newcommand{\e}{\mathrm{e}}
\newcommand{\dA}{\dot{A}}

\newcommand{\bI}{\boldsymbol{I}}
\newcommand{\bq}{\boldsymbol{q}}
\newcommand{\bg}{\boldsymbol{g}}
\newcommand{\bgamma}{\boldsymbol{\gamma}}
\newcommand{\bkappa}{\boldsymbol{\kappa}}
\newcommand{\btheta}{\boldsymbol{\theta}}
\newcommand{\bsigma}{\boldsymbol{\sigma}}
\newcommand{\brho}{\boldsymbol{\rho}}
\newcommand{\bxi}{\boldsymbol{\xi}}
\newcommand{\bF}{\boldsymbol{F}}
\newcommand{\bT}{\boldsymbol{T}}
\newcommand{\bS}{\boldsymbol{S}}
\newcommand{\bR}{\boldsymbol{R}}
\newcommand{\bQ}{\boldsymbol{Q}}
\newcommand{\bX}{\boldsymbol{X}}
\newcommand{\bK}{\boldsymbol{K}}

\newcommand{\bv}{\boldsymbol{v}}

\newcommand{\btau}{\boldsymbol{\tau}}
\newcommand{\bSigma}{\boldsymbol{\Sigma}}

\newcommand{\bomega}{\boldsymbol{\omega}}

\newcommand{\bPsi}{\boldsymbol{\Psi}}
\newcommand{\balpha}{\boldsymbol{\alpha}}
\newcommand{\bchi}{\boldsymbol{\chi}}

\newcommand{\bA}{{\boldsymbol{A}}}
\newcommand{\bL}{{\boldsymbol{L}}}
\newcommand{\blambda}{{\boldsymbol{\lambda}}}
\newcommand{\diff}{{\rm d}}
\newcommand{\hr}{\hat{r}}

\newcommand{\hht}{\hat{t}}

\newcommand{\hx}{\hat{x}}
\newcommand{\hs}{\hat{s}}
\newcommand{\hy}{\hat{y}}
\newcommand{\hz}{\hat{z}}

\newcommand{\hj}{\hat{j}}

\newcommand{\zq}{\tilde{q}}

\newcommand{\nn}{{\nonumber}}

\newcommand{\dynkinradius}{.05cm}
\newcommand{\dynkinstep}{.44cm}
\newcommand{\dynkindot}[2]{\draw (\dynkinstep*#1,\dynkinstep*#2) circle (\dynkinradius);}

\newcommand{\dynkinline}[4]{\draw[thin] (\dynkinstep*#1+\dynkinradius,\dynkinstep*#2) -- (\dynkinstep*#3-\dynkinradius,\dynkinstep*#4);}

\newcommand{\dynkindots}[4]{\draw[thick,dotted] (\dynkinstep*#1+\dynkinradius,\dynkinstep*#2) -- (\dynkinstep*#3-\dynkinradius,\dynkinstep*#4);}

\newenvironment{dynkin}{\begin{tikzpicture}[decoration={markings,mark=at position 0.7 with {\arrow{>}}}]}{\end{tikzpicture}}

\tikzset{
    vector/.style = {
        thick,
        > = stealth',
    },
    axis/.style = {
        very thin,
        > = stealth',
    },
}

\begin{document}

\markboth{Tomi Koivisto}
{Geometrical foundations of gravity}

%
\catchline{}{}{}{}{}
%

\title{On an integrable geometrical foundation of gravity}

\author{TOMI KOIVISTO}

\address{Nordita, KTH Royal Institute of Technology and Stockholm University, \\
Roslagstullsbacken 23, 10691 Stockholm, Sweden\\
\email{tomik@astro.uio.no} }

\maketitle

\begin{history}
\received{(Day Month Year)}
\revised{(Day Month Year)}
\end{history}

\begin{abstract}

In a talk at the conference {\it Geometrical Foundations of Gravity at Tartu 2017}, it was suggested that the affine spacetime connection could be associated with purely fictitious forces. 
This leads to gravitation in a flat and smooth geometry. Fermions are found to nevertheless couple with the metrical connection and a phase gauge field. 
The theory is reviewed in this proceeding in a Palatini, and in a metric-affine gauge formulation. 
 
\end{abstract}


\section{Introduction}	

Lie algebras are manifolds. In given a basis, a Lie algebra $\mathfrak{g}$ defines a differential-topological structure that is locally Euclidean. 
Besides a Euclidean intuition, a principle of relativity is inbuilt into the definition of a manifold, as we can locally map it into an infinite number of different
coordinate systems which are all diffeomorphically related to each other, but none of which is {\it a priori} preferred over the others \cite{weyl}. 

To consider, say a vector field $V^a$, one needs to set it in some {\it frame}, and to refer to it in the manifold, the ''formal scaffolding'' of some 
coordinate {\it labels} is needed. Basically,  a covector $\bbe_a$ sets up a frame, and once it is considered as a covector field $\bbe_a(x^\alpha)$, set up is a tangent frame bundle with the structure of the $\mathfrak{g}$ that transforms the field. 
In a representation furnished by some matrices $\Lambda^a{}_b$, such that $V^a \rightarrow \Lambda^a{}_b V^b$ for the vector, and $\bbe_a \rightarrow (\Lambda^{-1})^b{}_a \bbe_b$ for the frame, the invariance under these transformations reflects the arbitrariness in setting up the frame.
The same matrices can be used to carry out the diffeomorphisms, invariance under which reflects the arbitrariness of the coordinate labels, and this way
 the quotient of the infinite-dimensional diffeomorphism group and its linear subgroup is realized non-linearly over those matrices. Still, it would be a mistake to think of this as a Yang-Mills theory of the diffeomorphisms, for one reason the commutation relations of the latter have nothing to do with the commutation relations of the $\mathfrak{g}$ \cite{DeWitt:1965jb}. 
However, {\it symmetric teleparallelism} may offer a totally new perspective to this. 

The symmetric teleparallel geometry was presented in an ingenious paper of Nester and Yo \cite{Nester:1998mp}. The geometry is trivial, in the sense that there is no curvature nor torsion. Technically, calculations in coordinates are ''legitimized'' by bestowing covariance on them. 
This allows the interpretation that {\it in  coincidence} \cite{BeltranJimenez:2017tkd}, we may not only have undone the inertial (pseudo-)rotations of the frame, but also the inhomogeneous inertial effects which are the result of the displacements of the frame. Obviously, we would like to identify the coordinate and momenta that obey the canonical commutation relations, to begin a quantum mechanical discussion of spacetime and gravitation. As it is well known, General Relativity (GR) cannot separate gravitation from inertial effects, but miraculously that becomes possible by acknowledging the frame $\bbe_a$ \cite{Moller:1961jj}. The association of the pure-gauge spin connection with inertial effects has been clarified in depth in the context of teleparallel gravity \cite{Pereira}, 
where it is known e.g. that the suitable choice of the Lorentz reference frame eliminates the divergent boundary terms in the action \cite{Lucas:2009nq}.
Yet, besides the {\it flatness} of the connection, one may require its {\it smoothness}, in order to establish both the canonical frame and the canonical coordinates. 

This logic is supported by two very elementary findings in the resulting formulation of GR \cite{BeltranJimenez:2017tkd}:
\begin{itemize}
\item the integrable general linear connection is a translation and 
\item GR is the unique translation-invariant metric quadratic form. 
\end{itemize}
This may be explained better after introducing the connection below in Section \ref{geometry}.  We will consider paths on a manifold and deduce the gravitational force associated with 
the geodesic connection according to the equivalence principle. In the following Section \ref{sec:metrics} we take into account a metric and verify the second proposition above. In a brief digression to field theory, we then also comment on the spectrum of more general quadratic forms, and on the bootstrap of the unique invariant form. It is useful to review symmetric teleparallelism in the metric-affine gauge formalism \cite{Hehl:1994ue}. Especially, Adak {\it et al} used the language of differential
forms to develop the theory \cite{Adak:2011rr}, from which our Palatini formulation recovers consistently a holonomic aspect as will be verified in Section \ref{sec:vectors}. There we also suggest a semi-simple extension towards a more complete theory. In Section \ref{sec:spinors}, fermions are incorporated by postulating the Hermitian map between the 4-dimensional and a 2-dimensional (but complex) general linear transformation. The main result of that Section, which is not new either, is that spinors  decouple from non-metricity - except for an imaginary one-form which could, eventually, provide us the electromagnetic potential. 

To conclude in Section \ref{sec:conclusions}, we speculate on the unitary foundation of physics upon a {\it perfect geometry}, wherein the spacetime emerges as an integrable quotient.  

\section{Affine geometry}
\label{geometry}

\begin{center}
\begin{figure}
\begin{tikzpicture}[scale=0.8]

    \draw[axis,->][yshift=-1cm,xshift=0.5cm] (12,3,0) coordinate (O) -- (13,3,0) node[anchor=north east]{$$}; 
    \draw[axis,->][yshift=-1cm,xshift=0.5cm]  (12,3,0) -- (12,4,0) node[right,anchor=west]{$$}; 
    \draw[axis,->][yshift=-1cm,xshift=0.5cm]  (12,3,0) -- (12,3,1) node[anchor=south]{$$};
    \draw [gray, line width=1pt][yshift=-1cm,xshift=0.5cm]  plot [smooth] coordinates {(11.5,3.5) (12,3,0) (12.25,4) (12.5,3.5)(12.7,4.5) (13,3.5)};

  \draw[axis,->][yshift=2.5cm,xshift=0.5cm] (12,3,0) coordinate (O) -- (13,3,0) node[anchor=north east]{$$}; 
    \draw[axis,->][yshift=2.5cm,xshift=0.5cm]  (12,3,0) -- (12,4,0) node[right,anchor=west]{$$}; 
    \draw[axis,->][yshift=2.5cm,xshift=0.5cm]  (12,3,0) -- (12,3,1) node[anchor=south]{$$};
    \draw [darkgray, line width=1pt][yshift=2.5cm,xshift=0.5cm]   plot [smooth, tension=1] coordinates {(11,2.5) (13,3.5)};
    
	\draw[ultra thick] (0,5)--(2,5);
	\node at (1,5) {$\bullet$};
	\node at (1,5.35) {$0$};
	\node at (0.5,5) {$($};
	\node at (1.5,5) {$)$};
	\node at (0,5.35) {$\mathbb{C}$};
	\draw[<-][thick,color=gray] (0.75,4.8)--(0.75,4.2);
	\node at (0.35,4.5) {\tiny{$\textcolor{gray}{\mu^{-1}}$}};
	\draw[->][thick,color=gray] plot [smooth, tension=1.5] coordinates { (1.25,3.75) (3.5,3.5) (5.95,3.9)};
	\node at (3.15,3.2) {$\vec{\gamma}\circ{\textcolor{gray} \mu}$};
	\node at (2.15,2.8) {\tiny{\textcolor{gray}{another parameterization of} \textcolor{black}{the path}}};
	\draw[ultra thick] (0,4)--(2,4);
	\node at (1,4) {$\bullet$};
	\node at (1,3.65) {$0$};
	\node at (0.5,4) {$($};
	\node at (1.5,4) {$)$};
	\node at (0,3.65) {$\mathbb{C}$};
	\draw[->][thick,color=gray] (1.25,4.8)--(1.25,4.2);
	\node at (1.55,4.5) {\tiny{$ \textcolor{gray}{\mu}$}};
	\draw[->][thick,color=gray] plot [smooth, tension=1.5] coordinates { (1.25,5.25) (3.5,5.2) (5.95,4.1)};
	\node at (3.35,5.6) {$\textcolor{black}{\vec{\gamma}}$};
	\node at (3.35,6.0) {\tiny{\textcolor{gray}{a parameterization of} \textcolor{black}{the path}}};

	\draw[smooth cycle,tension=.7] plot coordinates{(4,3) (5.5,5) (7,5) (9,6) (9.5,3)};	
        \draw[ultra thick,black] plot [smooth, tension=1.5] coordinates { (5,3) (6,4) (7,3) (8,5)};
        \node (p) at (6,4) {$\bullet$};
        \node at (6.2,4.2) {$p$};
        \node at (9,3.5) {\tiny{manifold}};

         \draw[->][thick,color=gray] plot  [smooth, tension=1.5] coordinates { (10,3.5)  (11,2.7)};
	 \node at (11.25,2.4) {\textcolor{gray}{$y($}$\gamma$\textcolor{gray}{$)$}};
	  \draw[->][thick,color=darkgray] plot  [smooth, tension=1.5] coordinates { (10,4.5)  (11,5.25)};
	 \node at (11.25,6.0) {\textcolor{darkgray}{$x($}$\gamma$\textcolor{darkgray}{$)$}};
         \draw[->][thick,color=black] plot  [smooth, tension=1.5] coordinates { (12.6,4.6)  (12.6,3.6)};
         \draw[<-][thick,color=black] plot  [smooth, tension=1.5] coordinates { (12.3,4.6)  (12.3,3.6)};
          \node at (13.4,4.25) {\tiny{$y\circ x^{-1}$}};
          \node at (11.5,4.25) {\tiny{$x\circ y^{-1}$}};

  \end{tikzpicture}
\caption{A path is independent of its parameterization, and it has arbitrary coordinatizations. A real projection corresponds to a {\it scaling}, and an imaginary projection corresponds to a {\it phasing}. The gauge potential of the former will be introduced
as $\bQ$ in Section \ref{sec:vectors} and the gauge potential of the latter will appear as $i\bq$ in Section \ref{sec:spinors}. The semi-metric connection has the components (\ref{disformation2b}). 
\label{manifold}}
   \end{figure}
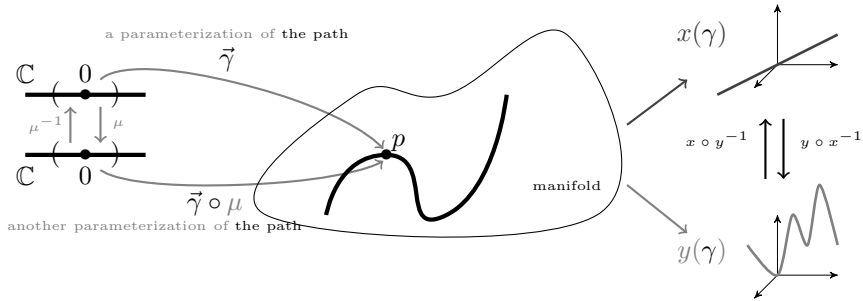
\end{center}

Let us draw a curve $\gamma(t): \mathbb{R} \rightarrow M$ on a manifold $M$, see the left hand side of Fig. \ref{manifold}. In the geometry of paths, the image of the curve
describes the structure of the {\it path} which is independent of its parameterization. This encapsulates an insight in the approach of Ehlers, Pirani and Schild which separates the causal ({\it i.e.} conformal) structure from the inertial ({\it i.e.} projective) structure, see \cite{weyl,Capozziello:2012eu}.
We then consider two coordinatizations for the curve $\gamma(t)$, $x^\mu$ and $y^\mu$, as in the right hand side of the Fig. \ref{manifold}. 
If we denote 
\be
\gamma^\mu(x)  = x^\mu\circ\gamma\,, \quad {\gamma}^\mu(y)=y^\mu\circ\gamma\,,
\ee
then it follows that the first derivatives transform linearly, so that we can write
\be
\dot{\gamma}^\mu(x)=\frac{\diff}{\diff t}\gamma^\mu(x)\,, \quad \dot{\gamma}^\mu(y)=\lp\frac{\partial y^\mu}{\partial x^\nu}\rp\dot{\gamma}^\nu(x)\,,
\ee
but the second derivatives have an inhomogeneous transformation property and we have to write
\be
\ddot{\gamma}^\mu(x)=\frac{\diff^2}{\diff t^2}\gamma^\mu(x)\,, \quad
   \ddot{\gamma}^\mu(y)=\lp\frac{\partial y^\mu}{\partial x^\nu}\rp\dot{\gamma}^\nu(x) + \lp\frac{\partial^2 y^\mu}{\partial x^\alpha x^\beta}\rp\dot{\gamma}^\alpha(x)\dot{\gamma}^\beta(x)\,.  
\ee
To obtain a covariant definition of the acceleration we have to therefore acknowledge the compensating field, $\Gamma^\mu$, which has exactly this same transformation law,
\be \label{gamma}
\Gamma^\mu(y) =  \lp\frac{\partial y^\mu}{\partial x^\nu}\rp\Gamma^\nu(x)+ \lp\frac{\partial^2 y^\mu}{\partial x^\alpha x^\beta}\rp\dot{\gamma}^\alpha(x)\dot{\gamma}^\beta(x) =  \Gamma^\mu{}_{\alpha\beta}\dot{\gamma}^\alpha(x)\dot{\gamma}^\beta(x)\,.
\ee
In the second equality we have written the $\Gamma^\mu$ in its usual three-index form. 
In the presence of this field, there now exists a definition of free motion: it occurs along the geodesics of the connection $\Gamma^\mu$, which are the paths $\gamma^\mu(x)$ obeying the equation 
$\dot{\gamma}^\mu\nabla_\mu\dot{\gamma}^\alpha=\dot{\gamma}^\mu\lp \partial_\mu + \Gamma_\mu\rp \dot{\gamma}^\alpha=0$.  Thus there exists a covariant definition of acceleration, and the
 new covariant field, called $F^\mu$, arises if we are forced to describe the possible deviation from the geodesics. 

The law of motion under the force, assuming a coupling constant $m$, must have the form $F^\mu(x) =  m(\ddot{x}^\mu-\Gamma^\mu(x))$. 
In the canonical three-dimensional case $\vec{\gamma}(t)=\vec{x}(t): \mathbb{R} \rightarrow M \simeq \mathbb{R}^3$, we recognise Newton's law. If we set 
$\Gamma^\mu(x)=0$, we are confined to an absolute coordinate system, a structure in Newton's original formulation not found from our $M$. On the other hand, in the 4-dimensional picture where $x^\mu$ is a Lorentz vector, the GR equation for a geodesic $\gamma^\mu=x^\mu$, which reads $\ddot{x}^\alpha + \Gamma^\alpha{}_{\mu\nu}\dot{x}^\mu\dot{x}^\nu = 0$, is obtained by setting $F^\mu(x)=0$ and using the proper time as the parameterization $\frac{\diff}{\diff t}=\dot{x}^\mu\partial_\mu$.
This is indeed in accordance with the integrable gauge interpretation of the equivalence principle. 
A force would be related to the derivatives of the connection (\ref{gamma}), which cannot be made disappear by the diffeomorphism $y\circ x$. 
In the conventional geometrical interpretation of GR they give rise to curvature, which however is not the property of an integrable connection.

The curvature and the torsion of a general affine connection (\ref{gamma}) are defined as 
\ba
{R}^\alpha_{\phantom{\alpha}\beta\mu\nu}  & = &  2\partial_{[\mu} \Gamma^\alpha_{\phantom{\alpha}\nu]\beta}
+ 2\Gamma^\alpha_{\phantom{\alpha}[\mu\lvert\lambda\rvert}\Gamma^\lambda_{\phantom{\lambda}\nu]\beta}\,, \label{curvature} \\
{T}^\alpha_{\phantom{\alpha}\mu\nu} & = & 2\Gamma^\alpha_{\phantom{\alpha}[\mu\nu]}\,, \label{torsion}
\ea
respectively. The affine connection has a priori 64 degrees of freedom. 
If without curvature, it can still contain the 16 free parameters that determine the general linear transformation $\Lambda^\alpha{}_\beta$ generated from $\mathfrak{gl}(4)$,
\be \label{flat}
\flat{\Gamma}{}^\alpha{}_{\mu\nu}=(\Lambda^{-1})^\alpha{}_{\beta}\partial_\mu \Lambda^\beta_{\phantom{\beta}\nu}\,.
\ee
If we further restrict torsion to zero, we fix the six more degrees of freedom (that can be shown to be associated to the fixing of a Lorentz frame), since the solution 
$\partial_{[\mu} (\Lambda^{-1})^\beta{}_{\nu]}=0$ is that $(\Lambda^{-1})^\beta{}_{\nu}=\partial_\nu V^\beta(x)$ for some four functions $V^\beta(x)$. 
These functions we can clearly consider as the new coordinates in the transformation (\ref{gamma}), which is seen to represent the pure-gauge form of
the $\mathfrak{gl}(4,\mathbb{R})$ connection, where the four $V^\alpha(x)$ appear as Goldstone bosons.
Evidently there emerges a structure resembling the tetrad, components of the frame field, since we may see now the $\Lambda^{\mu}{}_\nu$ as the components 
of a holonomic (co)frame field $\bbe^a$ such that $\e^a{}_\mu= \partial_\mu y^a$. Relations between massless Goldstone bosons and the fields of the usual tetrad formulation of GR have been
clarified in the context of nonlinear realizations \cite{Isham:1971dv,Tresguerres:2000qn}. The tetrad structure can elaborate the manifold into a soldered bundle, and it can induce a spacetime metric. We will introduce the frame formalism for gravitation in Section \ref{sec:vectors} and for matter in Section \ref{sec:spinors}.  

 
\section{Metric geometry}
\label{sec:metrics}

At this point we take into account a metric tensor $g_{\mu\nu}$. As suggested above in Section \ref{geometry}, such an object can spontaneously emerge from an integrable affine connection. 
In any case, having both these structures available, it is then possible to define the non-metricity tensor 
\be \label{nonm}
Q_{\alpha\mu\nu}=\nabla_\alpha g_{\mu\nu}\,,
\ee
which exhibits only the effects of the connection that it has on magnitudes. 
In accordance with our gauge interpretation of gravitation, we would like require the invariance of the theory under translations of the connection. It was shown in \cite{BeltranJimenez:2017tkd} that there is a unique second-derivative quadratic form $Q^2$ built from (\ref{nonm}), which turned out to be dynamically equivalent to GR. 
To introduce this $Q^2$, consider first the prototype non-metricity, given by the trace $Q_\alpha=g^{\mu\nu}Q_{\alpha\mu\nu}$. The case that $Q_{\alpha\mu\nu}=
\footnotesize{\frac{1}{4}}Q_\alpha g_{\mu\nu}$ is called semi-metric. Whereas generic non-metricity appears in the connection as\footnote{A metrical term is ''disformation'',  an affine term is ''distortion'' (into which with torsion though is included the ''contortion''), and a material term is ''deformation''.}
\be
L^\alpha_{\phantom{\alpha}\mu\nu}  =   \frac{1}{2} Q^{\alpha}_{\phantom{\alpha}\mu\nu} - Q_{(\mu\phantom{\alpha}\nu)}^{\phantom{(\mu}\alpha}\,, \label{disformation} 
\ee
in the semi-metric case it reduces to the scale connection 
\be
4W^{\alpha}_{\phantom{\alpha}\mu\nu}  =  \frac{1}{2}g_{\mu\nu}Q^\alpha-\delta^\alpha_{(\mu}Q_{\nu)}\,. \label{disformation2b}
\ee
The quadratic form that decouples the flattened (\ref{flat}) and smoothened connection is
\be \label{qdef} 
Q^2 = \frac{1}{2}Q_{\alpha\beta\gamma} \lp L^{\alpha\beta\gamma} - 4W^{\alpha\beta\gamma}\rp\,.
\ee
By effecting a translation of the $Q$-scalar by a Planck length $\sim 1/M_{Pl}$,  we obtain $Q^2 \rightarrow Q e^{-\Box/M^2_{Pl}}Q$,  a possible ghost-free ultra-violet completion of the theory \cite{Conroy:2017yln}. On the other hand, the tensor $P^{\alpha\mu\nu} \equiv (L-4W)^{\alpha\mu\nu}$ could be seen as the {\it field excitation} of $Q_{\alpha\mu\nu}$, and the proposed ultra-violet completion as a non-local modification of their constitutive relation\footnote{By the generalization of the teleparallel constitutive relation \cite{Hohmann:2017duq}, the theory of non-local gravity has been formulated, where infrared effects emerge and may simulate dark matter \cite{nonlocal}.}.

We may also consider more general constitutive relations. They can determined from an action principle $\int \diff^n x \sqrt{-g}f$, where $f$ is an invariant formed from the metric tensor and its derivatives. An $f$ defines the constitutive relation
\be \label{nmsuper}
P^\alpha{}_{\mu\nu}  \equiv \frac{\delta f}{\delta Q_\alpha{}^{\mu\nu}}\,,
\ee
(where by variation we mean just the partial derivative unless the action is higher order derivative). 
An ambiguity that arises in the definition of $f$, or equivalently, in the constitutive relation (\ref{nmsuper}),
can be addressed if we insist on the geometric preference for relations that trivialise the equation of motion for the connection, 
\be \label{eom}
\nabla_\mu\nabla_\nu \lp \sqrt{-g}P^{\mu\nu}{}_\alpha\rp = 0\,.
\ee  
This is not a kinematical, but a dynamical equation because it depends upon the action. Two variational methods have been considered recently that lead to this result. 
\begin{itemize}

\item The Palatini variation \cite{BeltranJimenez:2017tkd}. The variational degrees of freedom are the fields in the $f(g_{\mu\nu},\Gamma^\alpha{}_{\mu\nu})$, and two Lagrange multipliers. Our action would be then $S_Q  =   \int \diff^n x\lb\frac{1}{2}\sqrt{-g} f +  \lambda_\alpha^{\phantom{\alpha}\beta\mu\nu} R^\alpha_{\phantom{\alpha}\beta\mu\nu} + \lambda_\alpha^{\phantom{\alpha}\mu\nu}T^\alpha_{\phantom{\alpha}\mu\nu}\rb$.
The desired geometry is set with Lagrange multiplier (tensor densities), and a technical complication in the complete analysis is that one needs to obtain the solutions also for 
these fields. They turned
out to be essential in the teleparallel Palatini theory where they mediate the dynamics of the torsion\footnote{For analyses of the Lagrange multipliers in the context of Poincar\'e gauge theory, see \cite{Blagojevic:2000qs}, and for a metric-affine approach to teleparallelism, see \cite{Obukhov:2002tm}. The gauge symmetry of the multipliers  symmetric
teleparallelism is also known \cite{Adak:2011rr} (the multipliers are introduced as $\blambda^a$ and $\blambda^a{}_b$ in section \ref{sec:vectors}).
In passing we mention the $C$-theories \cite{Amendola:2010bk}, where the connection is set to be compatible with a rescaled metric $\hat{g}_{\mu\nu}=C{g}_{\mu\nu}$ with a Lagrange multiplier.
In the prototype $f(g^{\mu\nu}\hat{R}_{\mu\nu})$ models \cite{Capozziello:2015lza}, we then reproduce
the Palatini version in the limit $C=f'$ and the metric version in the limit $C=1$. We notice that these coincide in the case of the Hilbert action. There is however yet a simpler
$C$-theory, the $C=0$ theory. It could provide an alternative means to trivialise the affine geometry, and to covariantise the Einstein action.} \cite{BeltranJimenez:2017tkd}. Teleparallelism thus resolves the problem of non-propagating torsion \cite{Hammond:2002rm}. 
\newline
\item The inertial variation \cite{Golovnev:2017dox}, see also \cite{Hohmann:2018rwf}. The variational degrees of freedom are the $f(g_{\mu\nu},\Lambda^\alpha{}_\beta)$, where $\Lambda^\alpha{}_{\beta}$ is a St\"uckelberg field, restoring the available gauge freedom. In Weitzenb\"ock geometry it introduces an antisymmetric tensor \cite{Golovnev:2017dox,Hohmann:2018rwf}, and in the symmetric teleparallel geometry, a vector \cite{BeltranJimenez:2017tkd,Conroy:2017yln}. In this proceeding we call this vector $V^\alpha$. 
Our action could be written just as $S_Q  =   \frac{1}{2}\int \diff^n x\sqrt{-g} f(g_{\mu\nu},V^\alpha)$, since then the variations are {\it a priori} restricted to take place within the torsion-free and curvature-free geometry.  
\end{itemize}
Both methods also yield the metric field equation,
\be \label{geom}
\frac{1}{\sqrt{-g}}\nabla_\alpha \lp\sqrt{-g}P^\alpha{}_{\mu\nu}\rp - \frac{\partial f}{\partial g^{\mu\nu}} -  \frac{1}{2}f g_{\mu\nu}= T_{\mu\nu}\,.
\ee 
In general, hypermomentum could be included as a source. The conservation of the energy-momentum, defined for matter with a Lagrangian $\mathcal{L}_m$ in terms of the metric-compatible 
covariant derivative $\mathcal{D}_\mu$ as
\be \label{conservation}
T_{\mu\nu} = \frac{-2}{\sqrt{-g}}\frac{\delta \sqrt{-g}\mathcal{L}_m}{\delta g^{\mu\nu}}\,, \quad \mathcal{D}_\mu T^{\mu\nu} = 0\,,
\ee
can be shown to result from the above two dynamical equations (\ref{eom}) and (\ref{geom}) by a direct calculation, though it is less onerous to deduce it from the invariance of the action \cite{Koivisto:2005yk}. 

Let us then take a look at the spectrum of a more generic constitutive relation. Leaving out odd and higher derivative terms, the most general $f$ may depend upon the 
 five invariants 
\begin{subequations}
\label{letters}
\ba
A & \equiv & Q_{\alpha\mu\nu}Q^{\alpha\mu\nu}\,, \quad
B  \equiv Q_{\alpha\mu\nu}Q^{\mu\alpha\nu}\,, \quad \\
 C &  \equiv & Q_\alpha Q^\alpha , \quad
D  \equiv  \tilde{Q}_\alpha \tilde{Q}^\alpha\,, \quad
E  \equiv  \tilde{Q}_\alpha {Q}^\alpha\,,
\ea
\end{subequations}
where we have defined the projective trace $\tilde{Q}_\alpha=g^{\mu\nu}Q_{\mu\alpha\nu}$. 
For an $f=f(A,B,C,D,E)$ we obtain the non-metricity conjugate
\be
P_\alpha{}^{\mu\nu} = f_A Q_\alpha^{\phantom{\alpha}\mu\nu} 
 +  f_B Q^{(\mu\phantom{\alpha}\nu)}_{\phantom{\mu}\alpha} 
+ f_C g^{\mu\nu}Q_\alpha   \nn \\ 
 +  f_D\delta^{(\mu}_\alpha\tilde{Q}^{\nu)}  + f_E\delta^{(\mu}_\alpha {Q}^{\nu)}\,.
\ee
Consider the theory perturbatively. We perturb the metric $g_{\mu\nu} = \eta_{\mu\nu} + \delta g_{\mu\nu}$ and the inertial connection $\Lambda^\alpha{}_{\beta} = \delta^\alpha_\beta + \partial_\alpha \delta V^\beta$. It is then easy
to see that the non-metricity, at the linear order, is given as the partial derivative of the combination $h_{\mu\nu} = \delta g_{\mu\nu} +  2\delta V_{(\mu,\nu)}$, a ''St\"uckelbergised'' metric, which is
invariant under diffeomorphisms. In fact the action is, to second order in perturbations, 
\ba \label{h_action}
2\mathcal{L} = \frac{1}{2}a\partial_\alpha h_{\mu\nu} \partial^\alpha h^{\mu\nu}+b\partial_\alpha h_{\mu\nu}\partial^\mu h^{\alpha\nu} +c\partial_\mu h^\mu{}_\nu\partial^\nu h
+\frac{1}{2} d\partial_\alpha h \partial^\alpha h + \delta g_{\mu\nu} \tau^{\mu\nu}\,,
\ea
where the coefficients of the quadratic terms appear as $a=2f_A(0)$, $b=f_B(0)+f_D(0)$, $c=f_E(0)$ and $d=2 f_C(0)$. The kinetic term for the invariant combination $h_{\mu\nu}$ can be inverted to find out the propagating degrees of freedom. We can read the propagator $\Pi(h)$ from the standard results \cite{VanNieuwenhuizen:1973fi} in terms of the spin projector operators $P^{(s)}$ for the spin $s$ 
\be \label{prop}
k^2 \Pi(h) = \frac{1}{a}P^{(2)} + \frac{1}{a-b}P^{(1)} + q_0^{-2}\lb \alpha P^{(0)}
+ (a+3d) P^{(\bar{0})} - \sqrt{3}(c+d) P^{(\times)}\rb\,.
\ee
The shorthands used are $\alpha =a+2b+2c+d$ and $q^2_0=(a+3d)\alpha-3(c+d)^2$. As well-known\footnote{Even if the pole appears
positive, the 4-vector in this (pseudo)-orthogonal representation splits into two transverse 3-vectors with opposite kinetic terms unless $a+b=0$. Another remark is that from the $\Pi^{(2)}$ we substract an unphysical scalar to obtain the graviton. Finally, it is not clear
to us whether the scalar modes would be problematical  \cite{VanNieuwenhuizen:1973fi}.} this would propagate pathological
vector modes unless the invariance under transverse diffeomorphisms was restored by $a+b=0$. 
We can extract the metric propagator by integrating out the affine connection.
The linearization of its equation of motion (\ref{eom}) in terms of the gauge-invariant combination $h_{\mu\nu}$ is
\be
0  =  a \Box \partial^\nu h_{\nu\alpha}  +   b \lp \partial_\alpha\partial^\mu\partial^\nu h_{\mu\nu} + \Box\partial^\nu h_{\nu\alpha}\rp  
  +  c\lp \Box \partial_\alpha h + \partial_\alpha\partial^\mu\partial^\nu h_{\mu\nu}\rp +
 d \Box \partial_\alpha h\,. \label{cEoMp}
\ee
The solution, plugged back back into an effective action for $\delta g$, implies the metric propagator \cite{Conroy:2017yln}
\be
k^2 \Pi(\delta g) = \frac{1}{a}P^{(2)} + \frac{1}{\lp a+3d-3\frac{c+d}{\alpha}\rp}P^{(0)}\,.
\ee
This contains no transverse modes. Thus, as could have been guessed, the vector in (\ref{prop}) is due to the presence of the $V^\alpha$ in the affine connection.

To end this subsection, we briefly consider the self-coupling of the linear theory, as originally performed to GR by Deser \cite{Deser:1969wk}.
If we were to present the field theory of a rank-2 symmetric tensor $h_{\mu\nu}$ \cite{Feynman:1996kb}, and to supersede all 
absolute structures with $\mathfrak{g}$-structures, we would avoid a partial derivative, and would understand the symmetry of the tensor already to imply the symbol $\eta^{\mu\nu}$ of the (pseudo)-orthogonal algebra. We would have presumably arrived at the quadratic theory 
\be \label{mass}
L_0 =    - \frac{1}{2} \nabla_\alpha h_{\mu\nu} \lp \eta^{\alpha\beta}\eta^{\mu(\nu}\eta^{\rho)\sigma} 
+ 2\eta^{\alpha\sigma}\eta^{\nu(\rho}\eta^{\nu)\beta} \rp \nabla_\beta h_{\rho\sigma} + h_{\mu\nu}\tau^{\mu\nu}\,,
\ee
which is invariant under the independent transformations of the rank-2 tensor $h_{\mu\nu}$.
Sources $\tau^{\mu\nu}$ are included. Especially, the gravitational field is taken to couple to its own energy-momentum. This is given by the variation wrt the metric $\eta^{\mu\nu}$ because the symmetric teleparallel geometry is non-orthonormal (or in terms of the frame instead of the metric, anholonomic, see next Section \ref{sec:vectors}).  
 The prescription for a Lagrangian $L$ is   
\be \label{emt}
\tau_{\mu\nu} =- \frac{2}{\sqrt{-\eta}} \frac{\partial (\sqrt{-\eta}L)}{\partial\eta^{\mu\nu}}\,,
\ee
which was discussed in \cite{Padmanabhan:2004xk}. 
In the present case $L=L_0$ in (\ref{mass}), we obtain an expression of the form
\be
\tau^{\mu\nu} = K^{\mu\nu\kappa\lambda\alpha\beta\rho\sigma}(\nabla_\alpha h_{\kappa\lambda})(\nabla_\beta h_{\rho\sigma})\,, 
\ee
with the tensor $K$ constructed with only the $\eta^{\mu\nu}$. This formula should reproduce the Tolman's expression for the (pseudo-tensor of) gravitational energy-momentum at the first non-trivial order. 
The tensor is self-coupled with
the source $\tau^{\mu\nu}_0$ given by (\ref{emt}) where now $L=L_0$. We can then regard $L_1 = L_0 + \lambda h_{\mu\nu}\tau^{\mu\nu}_0$ as the second-order approximation to the theory. 
Again we have to take into account its self-coupling, and now the source term $\tau^{\mu\nu}_1$ is given by the variation (\ref{emt}) with $L=L_1$. We obtain the next approximation,
$L_2=L_1 + h_{\mu\nu}\tau^{\mu\nu}_1$,  then let it self-couple to get the $L_3$, and so on until $L_\infty$.  The computation should be formally the same as in Ref. \cite{Padmanabhan:2004xk}, and the result should be $L_\infty = Q^2$.
In particular, the kinetic term corresponds not to the Hilbert action but to the Einstein action, now in the covariantised form (\ref{qdef}). 
That had been pointed out in \cite{Padmanabhan:2004xk}. It is also known that by introduction of a reference metric the Einstein action {\it i.e.} the $\{\}\{\}$-action can be promoted into a covariant  $\Gamma\Gamma$-action \cite{Tomboulis:2017fim}. These results are now reached from a different geometric foundation. 
Firstly, our general linear first principle has promoted the partial to the gauge-covariant derivatives, and each
of the infinite steps of the bootstrap manifestly respects the symmetry. Secondly, the prescription (\ref{emt}) is a consequence of the symmetric teleparallelism. 

We notice also that there is a {\it doubled} symmetry of the linear action, and it remains in the non-perturbative result (since the $Q^2$ in the coincident gauge is invariant up to a boundary term).  This property is less transparent in GR, though the bimetric variational principle reveals that there is pair of (massless) Fierz-Pauli terms hiding in the
Hilbert action \cite{bimetric}. We make a further remark in Section \ref{sec:spinors} when equipped with some algebra.

\section{Metric-affine geometry}
\label{sec:vectors}

In this Section, an aim is to understand the symmetric teleparallel Palatini theory from an anholonomic perspective of a metric-affine gauge theory. We will arrive at three-form field equation (\ref{stg1}) and at the four-form equation (\ref{stg2}).

First recall algebra. Let $\hj_A$ be the elements of a basis of $\mathfrak{g}$. 
The expansion coefficients of their brackets are known as the structure constants $f^C_{\phantom{C}AB}$,
\be \label{lie}
[\hj_A,\hj_B] = f^C_{\phantom{C}AB}\hj_C\,,
\ee
and the structure is locally Euclidean, as desired, if it exhibits the antisymmetry and the Jacobi closure, 
\be \label{jacobi}
f^A_{\phantom{A}(BC)}=0\,, \quad f^A_{\phantom{A}[BC}f^D_{\phantom{D}E]A}=0\,.
\ee
The is metric defined as
\be \label{metric}
\eta_{AB} = f^C_{\phantom{C}AD}f^D_{\phantom{D}BC}\,,
\ee
and up to an overall constant, it is representation-independent and obtainable as the trace of the 
commutator.
The metric can be used also to lower and raise the indices of the structure constants, and by using the Jacobi (\ref{jacobi}), one verifies that the
$f_{ABC}=g_{AD}f^D{}_{BC}$ is totally antisymmetric $f_{ABC}=f_{[ABC]}$. If $\mathfrak{g}$ is semi-simple, $\det{g}\neq 0$, so an inverse $\eta^{AB}\eta_{BC}=\delta^A_C$ exists and one can recover the structure constants from the totally antisymmetric symbols. 

Meanwhile in geometry, the connection is given by the dual one-forms $\bgamma^A$ of the generators $\hj_A$,
and their algebra (\ref{lie}) is expressed for the connection as the Maurer-Cartan structure equations,
\be \label{mc}
\diff \bgamma^A = -\frac{1}{2}f^A_{\phantom{A}BC}\bgamma^B \wedge \bgamma^C\,. 
\ee 
Their {\it integrability}, which follows from the Poincar\'e's lemma $\diff^2=0$, is the Jacobi identity for the structure constants (\ref{jacobi}).
 A change of the connection results in curvature, just as the $F^\mu$ in Section \ref{geometry} arises to describe deviation from geodesics,
\be \label{gcurvature}
\bF^A = \diff \bgamma^A + \frac{1}{2}f^A_{\phantom{A}BC}\bgamma^B \wedge \bgamma^C\,,
\ee
and using now the $\diff^2=0$ is gives
 \be \label{gbianchi}
 \diff\bF^A + f^A_{\phantom{A}BC}\bgamma^B\wedge\bF^C =0\,,
 \ee 
which in geometry are known as the Bianchi identities. For a quotient construction, one perhaps should assume an ideal $\mathfrak{a} \in \mathfrak{g}$ such that the exponentiated group is normal. Nevertheless, we shall extract the generators of the base $M$ and of the frame transformations from the same $\mathfrak{g}$. 
We denote $\balpha^{ab}$ the subset of the one-form duals of the generators of $\mathfrak{a}$, and the rest as $\btheta^a$. They span, respectively, the
fibers and the cotangent spaces to $M$. Then we can write, for example,
\be \label{gcurv}
\bF^A = \frac{1}{2}F^A_{\phantom{A}ab}\btheta^a \wedge \btheta^b\,.
\ee 
As we require the connection to remain horizontal, that is independent of the lifting, the curvature does not include any of the fibre-spanning one-forms.  

For example, the Poincar\'e algebra $\mathfrak{iso}(3,1)$ of elementary particles is given by the following commutation relations:
\be
\label{cr}
[\hr_{ab},\hr_{cd}]    =   2\lp \eta_{d[a}\hr_{b]c} -  \eta_{c[a}\hr_{b]d}\rp\,, \quad
[\hht_a,\hr_{bc}]  =   2\eta_{a[b}\hht_{c]}\,, \quad [\hht_a,\hht_{b}]  = 0 \,,
\ee
for the six generators $\hr^a{}_b$ of (pseudo)-rotations, for which $\hr_{ab}=-\hr_{ba}$, and the four translation generators $\hht_a$.
Reading the structure constants from (\ref{cr}), we can obtain the kinematics of the Poincar\'e gauge theory \cite{Obukhov:2006gea}. 
In the general linear case, the shear generators $\hs^a{}_b$ will complement the commutations relations (\ref{cr}) with 
\begin{subequations}
\label{crb}
\ba
\lb \hs_{ab},\hs_{cd}\rb   & = &   2\lp \eta_{c(a}\hr_{b)d} + \eta_{d(a}\hr_{b)c} \rp\,, \\
\lb \hs_{ab},\hr_{cd}\rb    & = &   2\lp \eta_{c(a}\hs_{b)d}-\eta_{d(a}\hs_{b)c} \rp\,, \\
\lb\hht_a,\hs_{bc}\rb    & = &  2\eta_{a(b}\hht_{c)}\,.
\ea
\end{subequations}
Note that the separation into symmetric and antisymmetric pieces is possible only wrt some suitable orthogonal structure, such as the $\eta_{ab}$ which we have assumed, though not its constancy. 
With this in mind, the connection can be decomposed accordingly,
\be \label{cartan}
\bgamma = \btheta^a\hht_a + \frac{1}{2}\brho^{ab}\hr_{ab} - \frac{1}{2}\bsigma^{ab}\hs_{ab}\,,
\ee
where we have now introduced the three sets of connection one-forms $\btheta^a$, $\brho^{ab}=\brho^{[ab]}$ and $\bsigma^{ab}=\bsigma^{(ab)}$.
The connection on the fibers is generated by any linear transformation $\balpha$, and we have $\bgamma = \btheta + \balpha$. We call the components of the homogeneous curvature as 
\ba
\bT^a & = & \diff \btheta^a + \balpha^a_{\phantom{a}b}\wedge\btheta^b\,, \\
\bA^{ab} & = & \diff \balpha^{ab} + \balpha^a_{\phantom{a}c}\wedge\balpha^{cb}\,. \label{ccurvatures}
\ea
The Bianchi identities follow directly by the exterior derivatives, or using (\ref{gbianchi}), as
\begin{subequations}
\ba
\diff \bT^a & = &  \btheta^b\wedge \bA^a_{\phantom{a}b} - \balpha^a_{\phantom{a}b}\wedge\bT^b\,, \\
 \eta^{bc}\diff \bR^a{}_c & = &-2\brho_c^{\phantom{c}[a}\wedge \bR^{b]c} - 2\bsigma_c^{\phantom{c}[a}\wedge \bS^{b]c} \,, \\
\eta^{bc} \diff \bS^a{}_c & = & 2 \bsigma_c^{\phantom{c}(a}\wedge\bR^{b)c} + 2\brho_c^{\phantom{c}(a}\wedge\bS^{b)c}\,,
\ea 
\end{subequations}
where we have further made explicit the decomposition into the ''metric'' $\bR^{ab}$ and ''non-metric'' $\bS^{ab}$ curvatures. 
It is useful to rewrite these with the fibre-covariant exterior derivative $D = \diff + \balpha$. They then assume the form
\be
D  \bT^a   =     \btheta^b\wedge \bA^a_{\phantom{a}b}\,, \quad
D  \bR^{ab}   =   -2\bsigma^{c[a}\wedge\bA^{b]}_{\phantom{b]}c}\,, \quad
D  \bS^{ab}   =   2\bsigma^{c(a}\wedge\bA^{b)}_{\phantom{b)}c}\,.
\ee 
Note that by construction, $\bT^a=D\btheta^a$. 

For concreteness, we will have a look at the gauge transformations. Consider the transformation $\delta_\epsilon$ given by the parameters $\left\{\epsilon^a,\epsilon^{ab}\right\}$, 
\be \label{gauget}
 \hat{\epsilon} = \epsilon^a\hht_a + \frac{1}{2}\epsilon^{[ab]}\hr_{ab} + \frac{1}{2}\epsilon^{(ab)}\hs_{ab}\,.
\ee
For the gauge potentials we have the rule
\be
\delta_{\hat{\epsilon}} \bgamma^A = -\diff \epsilon^A + f^A_{\phantom{A}BC}\epsilon^B\bgamma^C\,,
\ee
from which we obtain using again the commutations (\ref{cr},\ref{crb}),
\begin{subequations}
\ba
\delta_{\hat{\epsilon}} \btheta^a & = & -\diff \epsilon^a + \epsilon^a_{\phantom{a}c}\btheta_{c} + \epsilon_c\bA^{ca}\,, \\
\delta_{\hat{\epsilon}}\bA^{a}{}_b & = & -\diff \epsilon^a{}_b + \epsilon^a_{\phantom{a}c}\bA^{c}{}_b + \epsilon_{bc}\bA^{ac}\,.
\ea
\end{subequations}
These can be decomposed as follows:
\begin{subequations}
\label{transformations}
\ba
\delta_{\hat{\epsilon}} \btheta^a & = & -\diff \epsilon^a + \epsilon^{[ac]}\btheta_{c} - \epsilon^{(ac)}\btheta_c
  - \epsilon^c\brho^a_{\phantom{a}c} + \epsilon^c\bsigma^a_{\phantom{a}c}\,, \label{trule} \\ 
\delta_{\hat{\epsilon}}\brho^{ab} & = & -\diff (\eta_{bc}\epsilon^{[ac]}) + \epsilon^{[ac]}\brho_c^{\phantom{c}b} + \epsilon^{[bc]}\brho^a_{\phantom{a}c}
+  \epsilon^{(ac)}\bsigma_c^{\phantom{c}b} - \epsilon^{(bc)}\bsigma_c^{\phantom{c}a} \,, \label{rrule} \\
\delta_{\hat{\epsilon}}\bsigma^{ab}  & = & -\diff(\eta_{bc}\epsilon^{(ac)}) +  \epsilon^{(ac)}\brho_c^{\phantom{c}b} - \epsilon^{(bc)}\brho^a_{\phantom{a}c}
 +    \epsilon^{[ac]}\bsigma_c^{\phantom{c}b} +  \epsilon^{[bc]}\bsigma_c^{\phantom{c}a}\,. \label{qrule} 
\ea
\end{subequations}
We note that when the potential $\bsigma$ vanishes, a shear transformation will not affect the spin connection. 
By plugging these transformations (\ref{transformations}) into the expressions for the 
curvatures (\ref{ccurvatures}), we can obtain the behaviour of the latter under the action generated by (\ref{gauget}).
The translations modify only the torsion,
\be
\label{ttrans}
\delta_{\hht} \bT^a = \epsilon^c \bF_c^{\phantom{c}a}\,, \quad \delta_{\hht} \bA^a{}_b = 0\,.
\ee 
We can verify the standard property of Lorentz transformations:
\be \label{ltrans}
\delta_{\hr}  \bT^a  =   \epsilon^a_{\phantom{a}b}\bT^b\,, \quad
\delta_{\hr}  \bR^{ab}  =  \epsilon^a_{\phantom{a}c}\bR^{cb} +  \epsilon^b_{\phantom{b}c}\bR^{ac}\,, \quad
\delta_{\hr}  \bS^{ab}  =  \epsilon^a_{\phantom{a}c}\bS^{cb} +  \epsilon^b_{\phantom{b}c}\bS^{ac}\,.
\ee
The non-Lorentzian transformation, in contrast, rotates the symmetric and antisymmetric curvatures into each other:
\be \label{lqrans}
\delta_{\hs} \bT^a  =   \epsilon^a_{\phantom{a}b}\bT^b\,, \quad
\delta_{\hs} \bR^{ab}  =   \epsilon^a_{\phantom{a}c}\bS^{cb} - \epsilon^b_{\phantom{b}c}\bS^{ca}\,, \quad
\delta_{\hs} \bS^{ab}  =  \epsilon^a_{\phantom{a}c}\bR^{cb} + \epsilon^b_{\phantom{b}c}\bR^{ca} \,.
\ee
The effect is non-trivial only for pure shear transformations. We can separate the rescalings, also known as {\it dilations}, for which $\epsilon^{ab}=2\delta^{ab}\epsilon^c_{\phantom{c}c}=2\delta^{ab}\epsilon$, and see that they leave both of the curvatures invariant. The torsion has the unit weight, but the curvatures have the zero weight under rescalings.


The integrable geometry will be characterised by $\bF^A=0$ and the Maurer-Cartan equations (\ref{mc}). Then the relations in the previous paragraph are all trivial, but we should yet take 
into account the more elaborate structure due to the presence of a metric. This naturally introduces the non-metricity one-form, 
\be \label{qform1}
D\eta_{ab} \equiv \bQ_{ab}\,.
\ee
Its Bianchi identity is: $D\bQ_{ab}=\bS_{ab}$. 
It is conventional to consider the spacetime metric through the projection of the frame fields as $g_{\mu\nu}\diff x^\mu \otimes\diff x^\mu = \eta_{ab}\bbe^a\otimes\bbe^b$.
However, we cannot identify the tetrad with the components of the translation potential one-form, as seen from (\ref{trule}).
We need to introduce some vector $V^a$, since only then $\ell \bbe^a = \btheta^a + D V^a$ 
has the correct transformation law (and with a scale $\ell$, it is dimensionless). Now a manifestation of the force appears in the torsion of the frame, $\ell D \bbe^a = \bT^a + \bA_b{}^a V^b$. If $\btheta^a=0$, the $V^a$ corresponds to the radius vector in Cartan geometry. 
The vector has a role of a Higgs field of translational symmetry, and it can be hidden into the translational piece of the connection in its nonlinear realization \cite{Tresguerres:2000qn}. Our interpretation is that the $V^\alpha$ is a translation of the integrable affinity that vanishes in coincidence \cite{BeltranJimenez:2017tkd}.  

Now with the metric and the tetrads at hand, the connection can be decomposed further. The
symmetric part splits into
\be \label{qform}
\bsigma_{ab} =  \frac{1}{2}\lp \diff \eta_{ab}- \bQ_{ab} \rp \,,
\ee
wherein a closed form may enter in a non-orthonormal frame.
The antisymmetric part contains the cotorsion defined via $\bK^a{}_b\wedge \bbe^a = \bT^a$, and the 
Levi-Civita one-form defined via $\bomega^a{}_b\wedge \bbe^b = -\diff \bbe^a$. One obtains \cite{Adak:2011rr}
\be \label{ab}
\brho_{ab} = -i_{[a}\diff \eta_{b]c}\bbe^c + \bomega_{ab} + \bK_{ab} +  i_{[a} \bQ_{b]c}\bbe^c\,. 
\ee
This generalises the more familiar GR spin connection $\bomega_{ab}$ to the general linear bundle. 
Note that by definition, $\bomega_{(ab)}=\bK_{(ab)}=0$, and these one-forms are introduced via the frame field both in the same fashion. 

We shall now specialise to the integrable connection, and can then set the contortion of the connection to vanish. To make contact with the Palatini formulation in Section \ref{sec:metrics}, 
we note that since (\ref{qform1}) is a tensor, we have simply $Q_{\alpha\mu\nu} = \e^a{}_\mu \e^b{}_\nu Q_{\alpha ab}$. Then we see the tensor (\ref{disformation}) appearing in the connection,
\be
L_{\alpha\mu\nu} =  \e^a{}_\mu \e^b{}_\nu \lp   i_{[a} \bQ_{b]c}\bbe^c  - \frac{1}{2} \bQ_{ab}  \rp(\partial_\alpha)\,.
\ee
Now recall that the components of the full affine connection are given as
\be \label{arel}
\Gamma^\alpha{}_{\mu\nu} = \e_a{}^\alpha D_\mu \e^a{}_\nu =  -\e^a{}_\nu D_\mu \e_a{}^\alpha\,.
\ee 
In the {\it orthogonal frame}, the one-form (\ref{qform}) is nothing but (minus twice) the shear gauge potential. We have then  
\be
\bQ_{ab}  \overset{\flat \perp}{=}  -2\bsigma_{ab}: \quad Q_{\alpha\mu\nu}  \overset{\flat \perp}{=} \eta_{ab}\partial_\alpha\lp \e^a{}_\mu e^b{}_\nu\rp - 2\Gamma_{(\mu\lvert\alpha\rvert\nu)}\,.
\ee
The connection can now vanish if the non-metricity is cancelled by the anholonomy (described by the Levi-Civita one-form $\bomega_{ab}$, often also written in terms of the Ricci rotation coefficients). Defining $\omega_{\alpha\beta\mu} = \e^a{}_\alpha\e^b{}_{\beta}\omega_{ab\mu}$, the affine connection can be expressed as
\be \label{g1}
\Gamma_{\alpha\mu\nu}  \overset{\flat \perp}{=}  \e_{a\alpha} \e^a{}_{\nu,\mu} + L_{\alpha\mu\nu} + \omega_{\alpha\nu\mu}\,.
\ee
In the {\it holonomic frame}, we have instead
\be
{\e}{}^a{}_\mu \overset{\flat h}{=} \delta^a_\mu: \quad Q_{\alpha\mu\nu}  \overset{\flat h}{=} \partial_\alpha \eta_{\mu\nu} - 2\Gamma_{(\mu\lvert\alpha\rvert\nu)}\,.
\ee
and the affine connection is also related more straightforwardly to $\balpha_{ab}$,
\be \label{g2}
\Gamma^{\alpha}{}_{\mu\nu}  \overset{\flat h}{=} L^{\alpha}{}_{\mu\nu} + \left\{^{\phantom{i} \alpha}_{\mu\nu}\right\}(\eta)\,,  
\ee  
where now appears the Christoffel symbol of the metric $\eta_{\mu\nu}$. 
The expressions (\ref{g1}) and (\ref{g2}) provide equivalent interpretations of the same affine spacetime geometry, 
the former through a non-trivial projection of an internal geometry, the latter through a faithful projection of a non-trivial internal geometry.   
Note that our definition of ''integrable'' is not the same as ''pure-gauge'' in the metric-affine theory, which regards also the non-metricity as a gauge field strength, see in 3.15 in \cite{Hehl:1994ue}. As noted there, the $\diff \eta_{ab} \neq 0$ in the holonomic frame can simulate non-metricity. In symmetric teleparallelism, which 
can accommodate non-metricity, the two can be made to cancel each other.
As is suggested by the construction (\ref{ab}), the anholonomy $\bomega_{ab}$ can simulate contorsion $\bK_{ab}$, but we see that  
the anholonomy in an orthogonal frame can simulate non-metricity as well and cancel its distortion of the connection.  


We now consider the action principle of a symmetric teleparallel theory. Our 5 scalar invariants (\ref{letters}) can be rewritten as
\ba
A  & = &   \langle \bQ_{ab}, \bQ^{ab} \rangle\,, \quad
B  =   \langle i_c\bQ_{ab},i^a \bQ^{bc} \rangle\,, \\
C  & = &  \eta^{ab}\eta^{cd}\langle \bQ_{ab}, \bQ_{cd} \rangle\,, \quad
D   =    i^b \bQ_{ab}, i_c \bQ^{ac}\,, \quad
E   =   \eta^{cd} i^b \bQ_{ab} i^a \bQ_{cd}\,.
\ea
Following Adak {\it et al} \cite{Adak:2011rr}, we consider the Lagrangian four-form involving two Lagrange multiplier two-forms, $\blambda^a$, $\blambda^a{}_b$, and a matter four-form $\bL_m$,
\be
\bL = \frac{1}{2} f(A,B,C,D,E)\ast 1 + \blambda_a\wedge \bT^a + \bA_a{}^b\wedge\blambda^a{}_b + \bL_m\,.
\ee
Define the variations
\be
\bSigma_a = \frac{\delta f}{\delta \bbe^a}\ast 1\,, \quad 
\bSigma^a{}_b = \frac{\delta f}{\delta \balpha_a{}^b}\ast 1\,, \quad 
\btau_a = \frac{\delta \bL_m}{\delta \bbe^a}\,, \quad 
\btau^a{}_b = \frac{\delta \bL_m}{\delta \balpha_a{}^b}\,. 
\ee
In terms of these three-forms, the field equations for the frame and the homogeneous potential are, respectively
\ba
\bSigma_a + D\blambda_a = \btau_a\,, \label{qdiff1} \\
\bSigma^a{}_b + \bbe^a\wedge\blambda_b + D\blambda^a{}_b = \btau^a{}_b\,.  \label{qdiff2}
\ea
The Lagrange multipliers set $\bT^a = \bA^a{}_b=0$. However, unlike in the Palatini formulation \cite{BeltranJimenez:2017tkd}, now they do not decouple from the dynamics.
For the first equation, we need the exterior derivative of $\blambda_a$. This can be deduced from the second equation as
\be
D \blambda_a = i_b D\lp \bSigma^b{}_a -  \btau^b{}_a\rp\,.
\ee
The exterior gauge-covariant derivative of the left hand side vanishes identically, which imposes an identity for the right hand side.
Thus, combining the above equation and its derivative, we can express the content of (\ref{qdiff1},\ref{qdiff2}) without the Lagrange multipliers:
\ba
i_b D \bSigma^b{}_a + \bSigma_a & = &  \btau_a + i_b D \btau^b{}_a\,, \label{stg1}\\ 
D i_b D \bSigma^b{}_a & = & D i_b D \btau^b{}_a\,. \label{stg2}
\ea
These are equivalent to our two equations (\ref{eom}) and (\ref{geom}) in the Palatini formulation (completed now including hypermomentum). 
Again, by the inertial variation one could deduce the same two equations. 

Last, recall algebra. As we wish to extract the translations from the $\mathfrak{g}$, it should have (at least) the rank 4.
Consider thus the linear group with an extra dimension.
The $\mathfrak{sl}(5)$ also has a non-degenerate Cartan-Killing form (\ref{metric}), so the metric as well as the connection is gotten from the algebra.
The translations are obtained from the rotations around the extra dimension. Since there is also shear, another set
of translations appear, resembling ''the special conformal transformation'' of the orthogonal case with 2 extra dimensions.  
For concreteness, we have in mind defining 
\be \label{basiscano}
\hx_a  =    \hr_{a 5} + \hs_{a 5}\,, \quad
\hy_a =  \hr_{a 5} - \hs_{a 5}\,, \quad \hz   = \frac{1}{2}\hs_{55}\,.
\ee
All the non-trivial commutation relations of the generators in this basis are:
\begin{subequations}
\label{catcano}
\begin{align}
\lb \hr_{ab},\hr_{cd}\rb  & =  2\lp \eta_{d[a}\hr_{b]c} -  \eta_{c[a}\hr_{b]d}\rp\,, \quad
&\lb \hs_{ab},\hs_{cd}\rb   & =  2\lp \eta_{c(a}\hr_{b)d} + \eta_{d(a}\hr_{b)c} \rp\,,  \\
\lb \hs_{ab},\hr_{cd}\rb   & =   2\lp \eta_{c(a}\zq_{b)d}-\eta_{d(a}\zq_{b)c} \rp\,,  \quad
&\lb \hx_a,\hr_{bc}\rb  & =   2\eta_{a[b}\hx_{c]} \,, \\
\lb \hy_a,\hr_{bc}\rb  & =   2\eta_{a[b}\hy_{c]} \,, \quad
&\lb \hx_a,\hs_{bc}\rb  & =   -2\eta_{a(b}\hx_{c)}\,, \\
\lb \hy_a,\hs_{bc}\rb  & =   2\eta_{a(b}\hy_{c)}\,, \quad
&{}[ \hz, \hx_a ] & =  -\hx_a\,, \label{cdil1}\\ 
{}[ \hz, \hy_a] & =  \hs_a\,, \label{cdil2}\quad
&{}[ \hx_a, \hy_b] & =  2( 2\eta_{ab}\hz-\hr_{ab}-\hs_{ab})\,.
\end{align}
\end{subequations}
With the two sets of translations, a quotient arises rather with 8 dimensions. Previously, an 8-dimensional quotient has been constructed
in a $\mathfrak{sl}(5)$ Yang-Mills theory  \cite{Assimos:2013eua} via an In\"on\"u-Wigner contraction, trivialising the extra dimensions. In the context of $\mathfrak{so}(4,2)$, a construction 
was called the {\it biconformal gauging} \cite{Wheeler:1997pc}. The flat biconformal bundle was shown to feature the symplectic structure of a phase space, and the trace one-form in a role similar to the electromagnetic field\footnote{The dilation curvature is of the form $\frac{1}{4}\diff \bQ + \bbe^a\wedge\bbie_a$, so flatness can be retained even when the one-form 
$\bQ=\eta^{ab} \bQ_{ab}$ has a non-trivial field strength. Lorentz force law and minimal coupling are recovered \cite{Wheeler:1997pc}.
It is also possible to employ the dilation-invariant metric $\bg = \bbe^a\oplus\bbie_a$ on a four-dimensional quotient as in the ''conformalised GR'' by Z\l o\'{s}nik  and Westman \cite{Zlosnik:2016fit}, see \cite{Rahmanpour:2018lzl} for a related discussion.} \cite{Wheeler:1997pc}.
The construction we propose is more straightforward than the previous ones in that it does not introduce the additional one-form duals to the $\hat{x}_a$ and $\hat{y}_b$. In the algebra (\ref{catcano}), these are already dual, $\langle \hat{x}^a, \hat{y}_b \rangle = \delta^a_b$. The generators of the co-tangent space can be directly related to the translation gauge potentials, and vice versa. This ''phase-spacetime'' will be presented elsewhere, but in the next Section \ref{sec:spinors} we will approach, with now a complex connection, the issue related to electromagnetism.

\section{Spinor geometry}
\label{sec:spinors}

Up to now we have considered the vector geodesic connection, but we know that matter fields are rather connected as spinors. Manifields are the infinite-dimensional representations of the linear group, world spinors being the holonomic manifields that represent the double cover of the group, see 4.1-4.7 in \cite{Hehl:1994ue}. However, in this Section we only aim to find the minimal gravitational coupling of fermions, and for this purpose consider a general linear transformation that is complex and 2-dimensional, in concert with the real and 4-dimensional general linear transformation. 



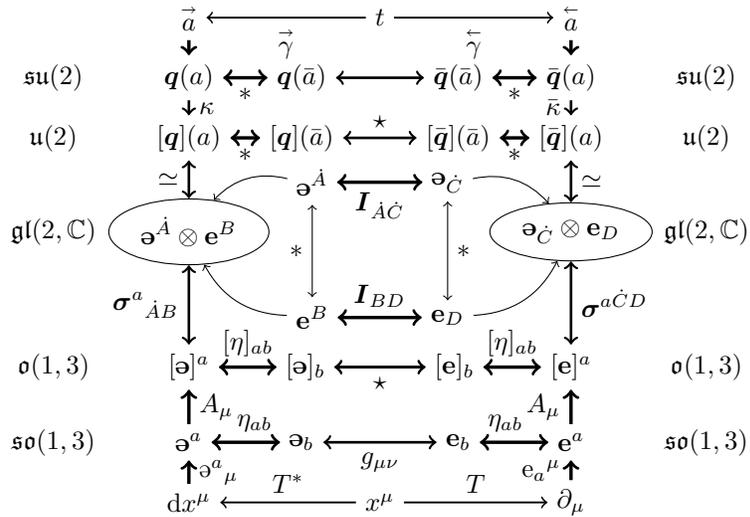
\begin{figure}
\begin{center}
\begin{tikzpicture}[node distance=1.8cm, auto]

  \node (S) {$\bbie^{\dA}$};
  \node (cS) [below of=S] {$\bbe^{B}$};
  \node (dS) [right of=S] {$\bbie_{{\dot{C}}}$};
   \node (dcS) [right of=cS] {$\bbe_{D}$};
  
  \draw[<->] (S) to node [left] {$\ast$}  (cS);
   \draw[<->] (dS) to node [right]  {$\ast$}  (dcS);
   
    \draw[<->][line width=0.45mm] (S) to node [below] {$\bI_{{\dA}\dot{C}}$} (dS);
     \draw[<->][line width=0.45mm] (cS) to node {$\bI_{BD}$} (dcS);
    
    \node (B) [node distance=1.14cm, xshift=-0.5cm,left of=cS, above of=cS] [ellipse,draw] {$\bbie^{\dot{A}}\otimes\bbe^{B}$};
 \node (dB) [node distance=1.14cm, xshift=0.5cm,  right of=dcS, above of=dcS] [ellipse,draw] {$ \bbie_{\dot{C}}\otimes\bbe_{D}$};
     
     \draw[->, bend right] (S) to (B);
     \draw[->, bend left] (cS) to (B);

    \draw[->, bend left] (dS) to (dB);
     \draw[->, bend right] (dcS) to   (dB);
     
     \node (T) [below of=B]  {$[\bbie]^a$};
     \node (dT) [below of=dB]  {$[\bbe]^a$};
     
       \draw[<->][line width=0.35mm]  (B) to node [left] {$\boldsymbol{\sigma}^a{}_{\dot{A}B}$} (T);
      
        \draw[<->][line width=0.35mm] (dB) to node [right] {$\boldsymbol{\sigma}^{a\dot{C}D}$} (dT);
   
   \node(apu1) [right of=T,node distance=1.55cm]  {$[\bbie]_b$};
    \node(apu2) [left of=dT,node distance=1.55cm]  {$[\bbe]_b$};
     \draw[<->][line width=0.45mm] (T) to node {$[\eta]_{ab}$} (apu1);
      \draw[<->][line width=0.45mm] (dT) to node [above] {$[\eta]_{ab}$} (apu2);
          \draw[<->][line width=0.35mm]  (apu1) to node [below] {\textcolor{black}{$\star$}}  (apu2);
           \draw[<->][line width=0.35mm]  (apu1) to node [above] {\textcolor{blue}{}}  (apu2);   
    \node (oo) [left of=T] {\textcolor{black}{$\mathfrak{o}(1,3)$}};
     \node (doo) [right of=dT] {\textcolor{black}{$\mathfrak{o}(1,3)$}};
    
   \node (Q) [above of=B,node distance=1.25cm]  {$[\bq](a)$};
   \node (dQ) [above of=dB,node distance=1.25cm]  {$[\bar{\bq}](a)$};
     \draw[<->][line width=0.35mm]  (B) to node [left] {\textcolor{black}{$\simeq$}} (Q);
     \draw[<->][line width=0.35mm] (dB) to node [right]  {\textcolor{black}{$\simeq$}}  (dQ);
      \node(apu3) [right of=Q,node distance=1.5cm]  {$[\bq](\bar{a})$};
    \node(apu4) [left of=dQ,node distance=1.5cm]  {$[\bar{\bq}](\bar{a})$};
     \draw[<->][line width=0.45mm] (Q) to node [below] {$\ast$} (apu3);
      \draw[<->][line width=0.45mm] (dQ) to node [below] {$\ast$} (apu4);
          \draw[<->][line width=0.35mm]  (apu3) to node [below] {\textcolor{blue}{}} node [above] {\textcolor{black}{$\star$}} (apu4);
          \node (uu) [left of=Q] {\textcolor{black}{$\mathfrak{u}(2)$}};
     \node (duu) [right of=dQ] {\textcolor{black}{$\mathfrak{u}(2)$}};

           \node at ($(B)!0.5!(dB)$)[yshift=-5]   {\textcolor{black}{}};
     
           \node (gg) [left of=B] {\textcolor{black}{$\mathfrak{gl}(2,\mathbb{C})$}};
     \node (dgg) [right of=dB] {\textcolor{black}{$\mathfrak{gl}(2,\mathbb{C})$}};

            \node (W) [below of=T,node distance=1.cm]  {$\bbie^a$};
            \node (dW) [below of=dT,node distance=1.cm]  {$\bbe^a$};
            \draw[->][line width=0.45mm] (W) to node [right] {$A_\mu$} (T);
            \draw[->][line width=0.45mm] (dW) to node [left] {$A_\mu$} (dT);
            \node(apu5) [right of=W,node distance=1.5cm]  {$\bbie_b$};
    \node(apu6) [left of=dW,node distance=1.5cm]  {$\bbe_b$};
     \draw[<->][line width=0.35mm]  (apu5) to node [below] {\textcolor{black}{$g_{\mu\nu}$}} node [above] {\textcolor{blue}{}}  (apu6);
       \node (ss) [left of=W] {\textcolor{black}{$\mathfrak{so}(1,3)$}};
     \node (dss) [right of=dW] {\textcolor{black}{$\mathfrak{so}(1,3)$}};

     \draw[<->][line width=0.45mm] (W) to node [above] {$\,\,\,\,\eta_{ab}$} (apu5);
      \draw[<->][line width=0.45mm] (dW) to node [above] {$\eta_{ab}\,\,\,\,$} (apu6);
           
            \node (X) [below of=W,node distance=0.8cm]  {$\diff x^\mu$};
            \node (dX) [below of=dW,node distance=0.8cm]  {$\partial_\mu$};
            \draw[->][line width=0.45mm] (X) to node [right] {$\ie^a{}_\mu$} (W);
            \draw[->][line width=0.45mm] (dX) to node [left] {$\e_a{}^\mu$} (dW);
            
            \node (M) at ($(X)!0.5!(dX)$) {$x^\mu$};
             \draw[->][line width=0.25mm] (M) to node [above] {$T^\ast$} (X);
             \draw[->][line width=0.25mm] (M) to node [above] {$T$} (dX);
             
            
              \node (U) [above of=Q,node distance=0.8cm]  {$\bq(a)$};
   \node (dU) [above of=dQ,node distance=0.8cm]  {$\bar{\bq}(a)$};
     \draw[<-][line width=0.35mm]  (Q) to node [right] {\textcolor{black}{$\kappa$}} (U);
     \draw[<-][line width=0.35mm] (dQ) to node [left]  {\textcolor{black}{$\bar{\kappa}$}}  (dU);
      \node(apu7) [right of=U,node distance=1.5cm]  {$\bq(\bar{a})$};
    \node(apu8) [left of=dU,node distance=1.5cm]  {$\bar{\bq}(\bar{a})$};
     \draw[<->][line width=0.45mm] (U) to node [below] {$\ast$} (apu7);
      \draw[<->][line width=0.45mm] (dU) to node [below] {$\ast$} (apu8);
          \draw[<->][line width=0.35mm]  (apu7) to node [below] {\textcolor{blue}{}} node [above] {\textcolor{black}{$$}} (apu8);
                     \node (su) [left of=U] {\textcolor{black}{$\mathfrak{su}(2)$}};
     \node (dsu) [right of=dU] {\textcolor{black}{$\mathfrak{su}(2)$}};

     \node (N) [above of=U,node distance=0.8cm]  {$\overset{\shortrightarrow}{a}$};
   \node (dN) [above of=dU,node distance=0.8cm]  {$\overset{\shortleftarrow}{a}$};
   \draw[<-][line width=0.45mm] (U) to node [above] {$$} (N);
             \draw[<-][line width=0.45mm] (dU) to node [above] {$$} (dN);
             \node (T) at ($(N)!0.5!(dN)$) {$t$};
             \draw[->][line width=0.25mm] (T) to node [below] {$\overset{\shortrightarrow}{\gamma}$} (N);
             \draw[->][line width=0.25mm] (T) to node [below] {$\overset{\shortleftarrow}{\gamma}$} (dN);
                      
\end{tikzpicture}
\caption{Relations of the various frames referred to in the text. The homothetic/orthogonal, the general/special linear, and the general/special unitary cases can be all mapped into each other, but there can be non-trivial global issues.   
\label{basises}}
\end{center}
\end{figure}

We set up a spinor frame $\bbe_A$  and a co-spinor frame 
  $\bbe^A$, where $A=1,2$. Different bases of the spinor and co-spinor frames are
  related by the $\Lambda^A{}_B$ generated from $\mathfrak{gl}(2,\mathbb{C})$, 
    \be
   \bbe^A  \rightarrow  \Lambda^A{}_B\bbe^B\,, \quad
   \bbe_A  \rightarrow  \bbe_B{(\Lambda^{-1})}{}^B{}_A\,.  \label{spinortrans}
  \ee
Since the bundle is complex, we can find there also the complex conjugates 
of each spinor and co-spinor space. Let us span the conjugate spinor bundle with the ''dotted spinor'' frame $\bbie_{\dot{A}}$ and the conjugate dual spinor bundle with the ''dotted co-spinor'' frame $\bbie^{\dot{A}}$. The transformations of these frames are as obvious as (\ref{spinortrans})
 We are then equipped with spinors with four types on indices on a principal fiber bundle with the structure of $\mathfrak{gl}(2,\mathbb{C})$.
Of special interest are the Hermitian second-rank spinors, due to to their celebrated isomorphism with vectors. 
The isomorphism can be given by the Hermitian map \cite{Hayashi:1976uz}
\be \label{sigmamap}
\bbe_a = \sigma_a^{\phantom{a}\dot{A}B}\bbie_{\dot{A}}\otimes\bbe_B\,.
\ee
Explicitly, we have for any vector $\bv=v^a\bbe_a = v^a\sigma_a{}^{\dot{A}B} \bbie_{\dot{A}}\otimes\bbe_{{B}}$, so that 
the spinor components of the vector are $v^{\dot{A}B}=v^a\sigma_a{}^{\dot{A}B}$. Similarly for a covector, $v_{\dot{A}B}=v_a\sigma^a{}_{\dot{A}B}$. 
The metric tensor of spacetime $\bg=g_{\mu\nu}\diff x^\mu\diff x^\nu$ is real and constructed from the complex spinor metric, $\bI$, as the direct product with the complex 
conjugate, $\bg = -\bar{\bI}\otimes\bI$. We arrive at a Penrose-Geroch spinorial deconstruction of Riemann's infinitesimal and relativistic generalization of the Pythagorean distance element:
\ba \label{spinormetric}
\bg   
= -\bar{\bI}\otimes\bI 
 =  -I_{\dot{A}\dot{C}}I_{BD}  {\bbie}^{\dot{A}}\otimes{\bbe}^B\otimes {\bbie}^{\dot{C}}\otimes{\bbe}^D\,.
\ea
Being the tangent space equipped with the metric $\eta_{ab}$, we can also display the relation 
\ba \label{spinormetric2}
\bg   =   \eta_{ab}\bbe^a\otimes \bbe^b
 =   \eta_{ab}\sigma^a_{\phantom{2}\dot{A}B}\sigma^b_{\phantom{2}\dot{C}D}  {\bbie}^{\dot{A}}\otimes{\bbe}^B\otimes {\bbie}^{\dot{C}}\otimes{\bbe}^D\,.
\ea
Fig. \ref{basises} illustrates some relationships between various frames on the manifold. On the unitary case, see  e.g. \cite{hyper,Bolokhov:2017ndw}.
At the end of this Section we consider the {\it hypermetric} extension where the elements in the above deconstructions could be picked from distinct bundles.

Now we should introduce the two-component spinor fields, $\bchi = \chi^A\bbe_A$ and 
$\bxi = \xi_{\dot{A}}\bbie^{\dot{A}}$, and form their direct sums
$\bPsi=\bchi\oplus\bar{\bxi}$ and $\bar{\bPsi} = \bxi\oplus\bar{\bchi}$,  
to arrange the pair into the 4-component Dirac spinor and its adjoint as
\be
\Psi = 
\lp
\begin{matrix}
 \chi_A  \\
 \xi^{\dot{A}}
\end{matrix} 
\rp\,,
\quad
\bar{\Psi} = 
\lp
\begin{matrix}
 \xi^{{A}}\,,\, \chi_{\dot{A}}
\end{matrix}
\rp\,.
\ee 
In the transformation (\ref{spinortrans}), the spinor components change obviously as 
\be
\xi^A   \rightarrow  \Lambda^A_{\phantom{A}B}\xi^B\,, \quad
\xi^{\dot{A}}  \rightarrow  \Lambda^{\dot{A}}_{\phantom{A}\dot{B}}\xi^{\dot{B}}\,, \quad
\chi_A  \rightarrow  \chi_B(\Lambda^{-1})^B_{\phantom{B}A}\,, \quad
\chi_{\dot{A}}    \rightarrow \chi_{\dot{B}} (\Lambda^{-1})^{\dot{B}}_{\phantom{B}\dot{A}}\,.
\ee
Let the spinor connection be $\overset{\ast}{\bomega}$. The covariant derivatives then act as 
\be \label{spinorders}
\overset{\ast}{\mathrm{D}} \xi^A   =  \diff \xi^A + \overset{\ast}{\bomega}{}^A_{\phantom{A}B}\xi^B\,,  \quad
\overset{\ast}{\mathrm{D}} \chi_{\dot{A}}  =  \diff\chi_{\dot{A}} + \overset{\ast}{\bomega}{}_{\dot{A}\phantom{B}}^{\phantom{ii}\dot{B}}\chi_{\dot{B}}\,.
\ee
We split the connection into the trace-free $\mathfrak{sl}(2,\mathbb{C})$ and the trace part as
\be
\overset{\ast}{\omega}{}^A_{\phantom{A}B\mu}  =  {\omega}{}^A_{\phantom{A}B\mu} + 
\frac{1}{2}\delta^A_B\lp \kappa + i q_\mu\rp\,, \quad
\overset{\ast}{\omega}{}^{\dot{A}}_{\phantom{A}\dot{B}\mu}  =  {\omega}{}^{\dot{A}}_{\phantom{A}\dot{B}\mu} + 
\frac{1}{2}\delta^{\dot{A}}_{\dot{B}}\lp \kappa_\mu + i q_\mu\rp\,. 
\ee
The spinor connection has 16 complex, or 32 real components, and the two one-forms $\bkappa$ and $\bq$ contain 8 of these.
 The relation of the spinor and the vector connections is determined by the compatibility with the mapping (\ref{sigmamap})
\ba
{\bomega}{}^A_{\phantom{A}B} & = & -\frac{1}{4}\sigma_a^{\phantom{a}\dot{C}A}
\sigma^b_{\phantom{b}\dot{C}B}\lp \balpha^a_{\phantom{a}b} - \frac{1}{4}\delta^a_b\eta^{cd} \bsigma_{cd}\rp\,, \\
{\bomega}{}_{\dot{A}\phantom{B}}^{\phantom{\dot{A}}\dot{B}}  & = &  -\frac{1}{4}\sigma^a_{\phantom{a}\dot{A}C}\sigma_{b\phantom{A}}^{\phantom{a}\dot{B}C}  
\lp \balpha^a_{\phantom{a}b} - \frac{1}{4}\delta^a_b\eta^{cd} \bsigma_{cd}\rp\,, \\ 
\bkappa & = & \frac{1}{2}\eta^{ab} \bsigma_{ab}\,. 
\ea 
The vector connection is therefore the mapping of the $\mathfrak{gl}(2,\mathbb{C})$ connection which loses the imaginary part of its trace. As noted in Ref. \cite{Hayashi:1976uz},
the one-form $\bq$ has nothing to do with the affine connection. 

We can now write down the Dirac Lagrangian with its linear kinetic term, 
\be
L = \frac{i}{2}\e_a^{\phantom{a}\mu}\lp \bar{\Psi}\gamma^a \mathrm{D}_\mu\Psi 
- ({\mathrm{D}}_\mu\bar{\Psi})\gamma^a\Psi\rp - m\bar{\Psi}\Psi\,.
\ee  
With our result for the relation of the connections, and some Clifford algebra, this becomes
\be
L  =  \frac{1}{2}\bar{\Psi}\lb (i \overset{\leftrightarrow}{\partial}_\mu + q_\mu) \gamma^\mu
- \frac{1}{3}\gamma^5\gamma_d\epsilon^{abcd}\alpha_{ab\mu}\e_{c}^{\phantom{c}\mu} - 2m\rb
\Psi\,.\label{l4}
\ee  
We confirm that the Dirac fermions couple only to the axial part of the spin connection, which generically consists of the rotation coefficients and the contorsion.
They are not affected by non-metricity\footnote{In the presence of non-metricity, the map
(\ref{sigmamap}) is not conserved: $\nabla_\alpha \sigma_\beta{}^{\dot{A}B}=L_{\alpha\beta}{}^\mu \sigma_\mu{}^{\dot{A}B}$. It seems to be possible to generalise this to
$\nabla_\alpha \sigma_\beta{}^{\dot{A}B}=(L_{\alpha\beta}{}^\mu + N_{\alpha\beta}{}^\mu)\sigma_\mu{}^{\dot{A}B}$, where $N_{\alpha(\beta\mu)} = 0$, but where 
$N_{\alpha[\beta\mu]} \neq 0$ would then appear in the Dirac equation \cite{Poberii:1998yq}.}. The field equations, for the spinor components rescaled by $\e^\frac{1}{2}$,
\begin{subequations} 
\label{spinoreom}
\ba
  \lb \gamma^\mu  (i{\partial}_\mu + \frac{1}{2}q_\mu) -  \frac{i}{2}\partial_\mu \gamma^\mu
 -  \frac{1}{3}\gamma^5\gamma_a\epsilon^{abcd}i_d \lp\bomega_{ab} + \bK_{ab} \rp - m\rb  \e^\frac{1}{2}\Psi\  & = & 0\,, \\
  \e^\frac{1}{2}\bar{\Psi} \lb \gamma^\mu  (i\overset{\leftarrow}{\partial}_\mu - \frac{1}{2}q_\mu) -  \frac{i}{2}\partial_\mu \gamma^\mu 
 +   \frac{1}{3}\gamma^5\gamma_a\epsilon^{abcd}i_d \lp\bomega_{ab} + \bK_{ab} \rp + m\rb & = & 0 \,,
\ea 
\end{subequations} 
imply the conservation of the probability current
\be
\partial_\mu j^\mu = 0\,, \quad j^\mu = \sqrt{-g}\bar{\Psi}\gamma^\mu\Psi\,. 
\ee
In symmetric teleparallel geometry, we can set the contorsion to vanish. Because the $\bQ_{ab}$ has totally decoupled, the fermions see only 
the anholonomy of the affine connection, $\bomega_{ab}$, and this part recovers the form of the standard minimally coupled Dirac theory in curved spacetime.  
The only non-metric interaction with the spacetime geometry is through the imaginary piece $i\bq$. Hayashi called this the {\it fermion-number gauge field}, because
its coupling is universal to all fermions \cite{Hayashi:1976uz}. In particular, though the $\bq$ enters precisely as the electromagnetic potential into the Dirac equation, 
there is no obvious electromagnetic interpretation for this phase coupling. 
However, Poberii has pointed out that one could enhance the symmetry with complex rescalings of the spinor metric,
to the effect that matter fields could be assigned with the desired conformal weights \cite{Poberii:1998yq}. 

We are contemplating the possibility that arises in the ''phase-spacetimes'' discussed at the end 
of the previous Section \ref{sec:vectors}. Underlying the bilateral frames $\bbe_a$ and $\bbie_a$, 
associated with the translation generators $\hx_a$ and $\hy_a$ in (\ref{basiscano}), respectively, there would be a double set of spinor frames as well\footnote{This gives 1-1 correspondence with the affine connection, which has double the number of independent components of the spinor connection. In fact only then can we arrange 
the ''concert'' in tune with the $\bsigma$-isomorphism, since the compatibility $\Lambda^a{}_b\sigma^{b \dot{A}}{}_{{B}} = \text{V}^{\dot{A}}{}_{\dot{C}}\sigma^{b \dot{C}}{}_{{D}} \Lambda^{D}{}_{B}$ allows to coordinate ${\text{V}^{-1}}{}^A{}_B = \Lambda^A{}_B$ within $\mathfrak{sl}(2,\mathbb{C})$ when the $\Lambda^a{}_b$ has the antisymmetric generators in $\mathfrak{so(3,1)}$ \cite{Hayashi:1976uz,Poberii:1998yq}, but the shear generators in $\mathfrak{sl(3,1)}$ call for the independence of $\text{V}^A{}_B$ and $\Lambda^A{}_D$ (while due to the dilation, we actually consider $[\eta_{ab}]$ not $\eta_{ab}$). }.
The semi-simplicity guarantees an invariant combination of the frames. For example the metric $\bg = \bbie^a\otimes\bbe_a$ is
invariant under the 5-dimensional linear transformation, but the metrics $\bg^+ = \bbe^a\otimes\bbe_a$ and $\bg^- = \bbie_a\otimes\bbie^a$ are not. The 
metric $\bg$ is ''neutral'' to non-metricity, whilst the latter two tensors come with the opposite non-metric ''charges''. The same recipe adapted to spinors yields two principal spinor frame bundles with opposite phasings, and various associated spinor frame bundles which could accommodate matter with different phase weights.  

Note that the bilateral frame structure is common to all $\mathfrak{g}$'s with (non-degenerate) metrics and of sufficiently high dimensions. 
The {\it double-copy} structure of the metric amplitudes \cite{Bern:2010ue} which is also generic \cite{Anastasiou:2017nsz} is derived 
 in an algebraic framework supporting a metric for the gauge and the gravity theory. When the scattering diagrams are suitably arranged by the antisymmetric and associative reasoning of the Lie algebra (\ref{jacobi}), the amplitudes can be matched by giving a double set of kinematical indices for the fluctuation of the spacetime metric in comparison to the gauge field fluctuation. This might reflect the double-diffeomorphism invariance of GR that became manifest in the $Q^2$-formulation of Section \ref{sec:metrics}. There we witnessed the covariant derivative $\nabla_\alpha$ of the improved field theory systematically copying the translational symmetries of the kinetic terms.

\section{Conclusions and perspectives}
\label{sec:conclusions}

The equivalence principle states that gravitation is indistinguishable from acceleration, and thus its effects can be locally transformed away by a diffeomorphic change of coordinates \cite{Feynman:1996kb}. In our interpretation, the former suggests that the gravitational connection is a translation, and the latter suggests that the inertial connection is integrable. This can be realised in the symmetric teleparallel geometry, as seen in Section \ref{geometry}. 
We considered the Palatini theory of gravity in this geometry in Section \ref{sec:metrics}, presenting the field equations (\ref{eom}) and (\ref{geom}). In Section \ref{sec:vectors} we in turn formulated a symmetric teleparallel metric-affine gauge theory of gravitation, and derived its field equations (\ref{stg1}) and (\ref{stg2}). 
 
With an integrable gravitational connection, a concern arises about the coupling of matter. Naively inserting the covariant derivatives of the gravitational connection into action of matter fields would result in their trivial(izable) geodesic trajectories, or add hypermomentum, both contradicting the usual conservation law (\ref{conservation}). This urged us to inquire the
coupling of fermions. In Section \ref{sec:spinors} we took into account a complex connection in an internal spinor space, and studied its relation to the affine connection in the external spacetime.
We found that the spinor connection is oblivious to the non-metricity in the affine connection, but on the other hand, possesses an independent piece of an imaginary non-metric potential. The gravitational coupling of the spinors thus turned out to be, in contrast to the naive expectation, equivalent to GR. 
Yet, the semi-Hermitian connection with the imaginary piece might link the electromagnetic phase to spacetime geometry, see however \cite{weyl,Hayashi:1976uz,Poberii:1998yq,Mannheim:2014ypa,Scholz:2012ev}.

As a repercussion of the Dirac equation, all matter at all scales exhibits chirality. 
It is natural that the spacetime is also a number, namely, the quaternion  of the unitary $\mathfrak{su}(2)$, rather than the 4-vector of the pseudo-orthogonal $\mathfrak{so}(3,1)$. 
The {\it unitary spacetime} would be non-vacuous! The cosmological constant would be topologically excluded. 
This is a prediction of the Conformal Algebrodynamic Theory, where we regard the external reality as an integrable quotient in a finite structure that is intelligible as a semi-simple hypercomplex computation. 
Only recently were discovered the four normed division algebras in mathematics, and the current standard models of the four interactions in physics, but their essence and unity was known to the ancients.
 
 As Weyl explained, the elementary axiomatic grounding of geometry leads to the abstract number concept \cite{weyl2012levels}.
 In the case of plane projective geometry, the incidence axioms alone lead to a number field, whose elements are dilations. Points and lines are ratios and triples of such numbers that obey the incidence equation. An interesting perspective to the path and its parameterizations in Fig. 1 is perhaps that numbers are not subject to size relations in algebra. For the geometric numbers to coincide with the continuum of ordinary reals, the axioms of order and continuity would need to be invoked.




\section*{Acknowledgments}

I am grateful to the organisers and participants of the extremely interesting conference Geometrical Foundations of Gravity
2017 in Tartu. Collaborations on the topic with Jose Beltr\'an Jim\'enez, Aindri\'u Conroy, Lavinia Heisenberg and Frank K\"onnig are acknowledged with great pleasure. I also would like to thank Prof. Ong for pointing out the reference \cite{Nester:1998mp} at the conclusion of the Nordita program Extended Theories of Gravity, and 
 I would like to thank Prof. Pavlov for giving the reference \cite{hyper} at FERT-2008. Needless to say, this proceeding represents solely its author's views. 

\appendix



\end{document}